\documentclass{article}

\usepackage[nonatbib, final]{tackling_climate_workshop_style}

\usepackage[utf8]{inputenc} 
\usepackage[T1]{fontenc}    
\usepackage[hidelinks]{hyperref}       
\usepackage{url}            
\usepackage{booktabs}       
\usepackage{amsfonts}       
\usepackage{nicefrac}       
\usepackage{microtype}      

\usepackage{mathtools}
\usepackage{xcolor}
\usepackage{subfigure}
\usepackage[capitalise, noabbrev]{cleveref}

\usepackage[doi=false,isbn=false,url=false,eprint=false,sorting=none,maxbibnames=30]{biblatex}
\usepackage[final]{changes}
\definechangesauthor[name={Ma\"elys}, color=blue]{MS}

\newcommand{\indep}{\perp\!\!\!\!\perp} 
\definecolor{purple}{HTML}{9673A6}
\definecolor{red}{HTML}{B85450}
\definecolor{blue}{HTML}{6C8EBF}
\definecolor{orange}{HTML}{D79B00}
\definecolor{green}{HTML}{82B366}
\definecolor{grey}{HTML}{666666}

\title{Using uncertainty-aware machine learning models to study aerosol-cloud interactions}

\author{%
  Ma\"elys Solal\\
  Department of Computer Science\\
  University of Oxford, UK\\
  \texttt{maelys.solal@ens.psl.eu}\\
  \And
  Andrew Jesson\\
  OATML\\
  Department of Computer Science\\
  University of Oxford, UK\\
  \And
  Yarin Gal\\
  OATML\\
  Department of Computer Science\\
  University of Oxford, UK\\
  \And
  Alyson Douglas\\
  AOPP\\
  Department of Physics\\
  University of Oxford, UK\\
}

\begin{document}

\maketitle

\begin{abstract}

Aerosol-cloud interactions (ACI) include various effects that result from aerosols entering a cloud, and affecting cloud properties. In general, an increase in aerosol concentration results in smaller droplet sizes which leads to larger, brighter, longer-lasting clouds that reflect more sunlight and cool the Earth. The strength of the effect is however heterogeneous, meaning it depends on the surrounding environment, making ACI one of the most uncertain effects in our current climate models. In our work, we use causal machine learning to estimate ACI from satellite observations by reframing the problem as a treatment (aerosol) and outcome (change in droplet radius). We predict the causal effect of aerosol on clouds with uncertainty bounds depending on the unknown factors that may be influencing the impact of aerosol. Of the three climate models evaluated, we find that only one plausibly recreates the trend, lending more credence to its estimate cooling due to ACI. 

\end{abstract}

\section{Introduction}

\begin{figure}[h]
    \centering
    \subfigure[Simple Causal Graph of ACI]{
    \includegraphics[width=.33\linewidth]{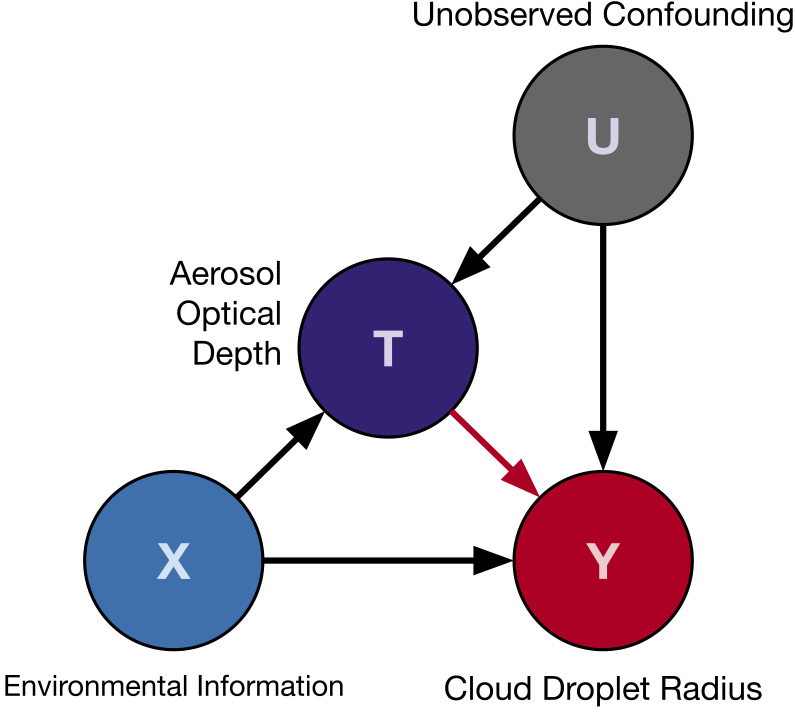}
    \label{fig:causal}
    }
    \subfigure[The Southeast Pacific]{
    \includegraphics[width=.3\linewidth]{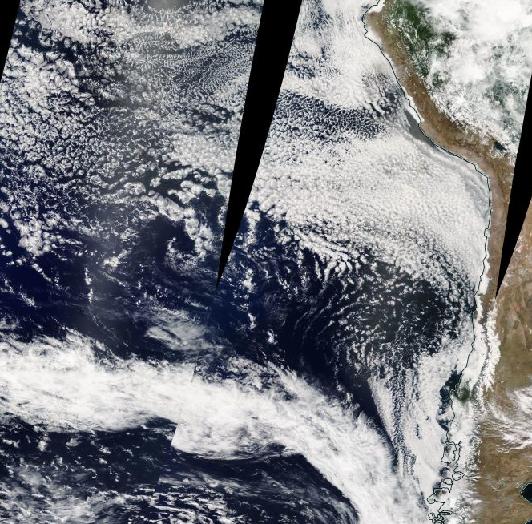}
    \label{fig:SEP}
    }
    \subfigure[The South Atlantic]{
    \includegraphics[width=.3\linewidth]{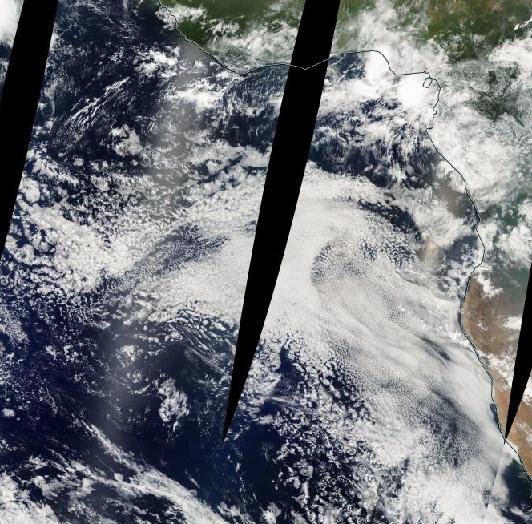}
    \label{fig:SA}
    }
    \caption{The causal graph underlining our knowledge of ACI and satellite imagery of the two regions analyzed, chosen due to their unique aerosol-cloud interactions and breadth of past studies to pull knowledge from.}
    \label{fig:causal_regions}
\end{figure}

Aerosol, in the form of pollution from human emissions, enters the atmosphere and eventually interacts with a cloud leading to aerosol-cloud interactions (ACI).
As aerosol enters the cloud, a causal chain of events catalyzes. 
It begins with aerosol particles activating as cloud droplet nuclei, which increases the number of droplets within the cloud, reducing the mean radius of cloud droplets to redistribute the water vapor, and eventually increasing the cloud's brightness (\cref{fig:causal}) \cite{twomey1984assessment}.
\replaced[id=MS]{Overall, an increase in atmospheric aerosol leads to larger, brighter, longer-lasting clouds that reflect more incoming sunlight.}{An increase in brightness leads to the cloud reflecting more incoming sunlight.}
ACI are thus a net cooling process and offset some fraction of warming due to rising levels of CO\(_{2}\).
\replaced[id=MS]{The strength of the effect is however dependent on the local environment surrounding the cloud. ACI remain one of the most uncertain effects in our current climate models, as current models are limited in their ability to simulate ACI with such environmental heterogeneity \cite{ipcc-ar5-wg1-ch7,ipcc-ar6-wg1-spm}.}{ACI are one of the most uncertain effects in our current climate models due in part to the strength of their effects being dependent on the local environment surrounding the cloud. Our current climate models are limited in their ability to simulate ACI with such environmental heterogeneity \cite{ipcc-ar5-wg1-ch7,ipcc-ar6-wg1-spm}.}
Climate models can only approximate ACI given their low spatial resolution and limited parameterizations, often dependent on only a few environmental parameters, such as the relative humidity within a grid cell.
These factors lead to increased uncertainty in future projections.
Currently, state-of-the-art climate models estimate that the range of cooling due to ACI may offset 0\%-50\% of the warming due to greenhouse gas emissions.

\replaced[id=MS]{This work uses causal machine learning to estimate ACI from satellite observations, by reframing the problem as a treatment (aerosol) and outcome (change in droplet radius). We predict the causal effect of aerosol on clouds and provide uncertainty bounds that we compare to the parameterizations of climate model ACI.}{This work uses causal machine learning to estimate ACI from satellite observations. We provide uncertainty bounds on these estimates and compare them to the parameterizations of climate model ACI.} 
\replaced[id=MS]{We consider uncertainty arising from violations of two assumptions: positivity (or overlap) and unconfoundedness (or no hidden confounding). Positivity violations are due to insufficient representation within the data for all treatment levels, such as "treating" cloud with aerosol. Unmeasured confounding are unobserved factors which influence both the treatment and outcomes, such as humidity causing aerosol swelling and altering cloud properties.}{Uncertainty in causal effect estimates can arise from violations of the positivity assumption, insufficient representation within the data for all treatment levels (such as "treating" cloud with aerosol), and from unmeasured confounding, when unobserved factors influence both the treatment and outcomes (such as humidity causing aerosol swelling and altering cloud properties).}
To better understand these individual sources of uncertainty, we use Overcast \cite{jesson2022scalable}, a prime example of the needs of a community such as ACI leading to methodological contributions in machine learning.
\deleted[id=MS]{Overcast estimates the effects of a continuous valued process and can account for the influence of heterogeneity, positivity violations, and confounding on the outcome.
In the real world, rarely are processes either `on' or `off', therefore creating a model that can encompass a range of effects depending on the magnitude of the treatment is vital to accurately model these processes.}
Compared to prior work such as \cite{jesson2021using}, we consider aerosol optical depth (AOD), our proxy for aerosol concentration, as a continuous treatment rather than discrete and perform an uncertainty-aware sensitivity analysis to study the consequences of possible violations of positivity and unconfoundedness.

\section{Methods}

\added[id=MS]{Following \cite{jesson2022scalable},} we use the potential outcomes framework to estimate the effect of a continuous treatment $\mathrm{T} \in \mathcal{T}$ (aerosol), on outcomes of interest $\mathrm{Y} \in \mathcal{Y}$ (cloud property), for a unit described by covariates $\mathbf{X} \in \mathcal{X}$ (environmental information)  as shown in \cref{fig:causal} \cite{rubinEstimatingcausaleffects1974, rubinRandomizationAnalysisExperimental1980, sekhonNeymanRubinModel2009, splawa-neymanApplicationProbabilityTheory1990}.
We call a potential outcome and denote by $\mathrm{Y}_{\mathrm{t}}$ what the outcome would have been if the treatment were $\mathrm{t}$.
\added[id=MS]{The covariates considered are relative humidity at 900, 850 and 700 millibar, sea surface temperature, vertical motion at 500 millibars, lower tropospheric stability, and effective inversion strength. The treatment is aerosol optical depth (AOD), a proxy for aerosol concentration. The outcome considered is the cloud droplet size ($r_e$).}
To estimate the treatment-effect, we study the conditional average potential outcome (CAPO) and the average potential outcome (APO)
\[
    \text{CAPO} = \mu(\mathbf{x}, \mathrm{t}) \coloneqq \mathbb{E}\left[\mathrm{Y}_{\mathrm{t}} \mid \mathbf{X} = \mathbf{x}\right],
    \quad \text{and} \quad
    \text{APO} = \mu(\mathrm{t}) \coloneqq \mathbb{E}\left[\mu(\mathbf{X}, \mathrm{t})\right],
\]
which can be identified from the observational distribution $P(\mathbf{X}, \mathrm{T}, \mathrm{Y})$ using 
\[
    \tilde{\mu}(\mathbf{x}, \mathrm{t}) \coloneqq \mathbb{E}\left[\mathrm{Y} \mid \mathrm{T}=\mathrm{t}, \mathbf{X} = \mathbf{x}\right] 
    \quad \text{and} \quad 
    \tilde{\mu}(\mathrm{t}) \coloneqq \mathbb{E}\left[\tilde{\mu}(\mathbf{X}, \mathrm{t})\right],
\]
and further assumptions (unconfoundedness, positivity, no-interference and consistency).
\deleted[id=MS]{Identifying the CAPO using finite observational satellite data depends on assumptions untestable from the data and challenged by the continuous treatment regime.}
\replaced[id=MS]{Here, we study the robustness of treatment-effect estimates to positivity and unconfoundedness violations (see \cref{app:assumptions} for more detail).}{Here, we study the robustness of treatment-effect estimates to violations of the unconfoundedness and the positivity assumptions.}
We compute uncertainty bounds corresponding to user-specified relaxations of these assumptions. 
The parameter $\Lambda$, for example, is set by the user to explain an assumed level of unmeasured confounding \cite{jesson2022scalable, jessonQuantifyingIgnoranceIndividualLevel2021, jessonIdentifyingcausaleffectinference2020}.
Some confounding influences are impossible to measure directly with satellites, such as humidity causing aerosol swelling and altering cloud properties, and the parameter $\Lambda$ can be used to encode an expert's belief in the influence of such confounders.
\deleted[id=MS]{Moreover, we rely on a proxy for aerosol, the aerosol optical depth (AOD), which has consequences distinct from both unmeasured confounding and positivity.}

We use daily mean, 1\(^{\circ}\) x 1\(^{\circ}\) of satellite observations in order to homogenize the data from the southeast Pacific and south Atlantic (Figures \ref{fig:SEP} and \ref{fig:SA}).
Mean droplet radius ($r_e$) from the MODIS instrument is used as our outcome for all experiments shown within. 
We employ aerosol optical depth from MERRA-2 to approximate the concentration of aerosol. 
Our environmental confounders are the relative humidity at 900, 850 and 700 millibars, the stability of the atmosphere, the sea surface temperature, and the vertical motion at 500 mb, all also from MERRA-2. \added[id=MS]{For more detail about data and implementation, please refer to \cref{app:data} and \cref{app:implementation}.}

\deleted[id=MS]{The models are neural-network architectures with two components: a feature extractor $\phi(\mathrm{x}; \mathbf{\theta})$ and a density estimator $f(\phi, \mathrm{t}; {\theta})$, \added[id=MS]{represented in \cref{app:model-arch}.}
The covariates $\mathbf{x}$ are given as input to the feature extractor, whose output is concatenated with the treatment $\mathrm{t}$ and given as input to the density estimator which outputs a Gaussian mixture density, $p(\mathrm{y} \mid \mathrm{t}, \mathbf{x}, {\theta})$, from which we can sample.
The feature extractor uses attention mechanisms to model the spatio-temporal correlations between the covariates on a given day using the geographical coordinates of the observations.
\added[id=MS]{Further implementation detail are given in \cref{app:implementation}.}}

\section{Results}

\subsection{Deducing reasonable treatment-effect bounds using domain knowledge}
 
Unlike past studies which only crudely estimate an uncertainty range due to quantifiable effects, we are able to derive confidence intervals dependent on the influence of confounding by varying $\Lambda$.
Since it is impossible to know the strength of the confounding effect from observed data alone, we propose a method to select a reasonable $\Lambda$ by contrasting two geographical regions. 
We contrast the South-East Pacific and the South Atlantic because these regions have different environmental confounders of ACI, for example aerosol type, aerosol hygroscopicity, aerosol size.
These are important confounders, but are unfortunately not included in the available data.
So we select the parameter $\Lambda$ for the Pacific region such that the treatment effect bounds for the Pacific region cover the effect bounds for the Atlantic region under the assumption of no hidden confounding ($\Lambda \to 1$) for the Atlantic region, as shown in \cref{fig:lambda}.
Setting $\Lambda$  to 1.07 gives bounds for the pacific region that reasonably account for the potential bias induced by the unmodeled confounders.
While a larger $\Lambda$ could still be sensible due to other drivers of confounding, domain knowledge informs us that these are the main missing physical mechanisms.

\subsection{Evaluating climate models using machine learning}

\begin{figure}[htbp]
    \centering
    \subfigure[Choosing $\Lambda$ using two geographical regions]{
    \includegraphics[width=.4\linewidth]{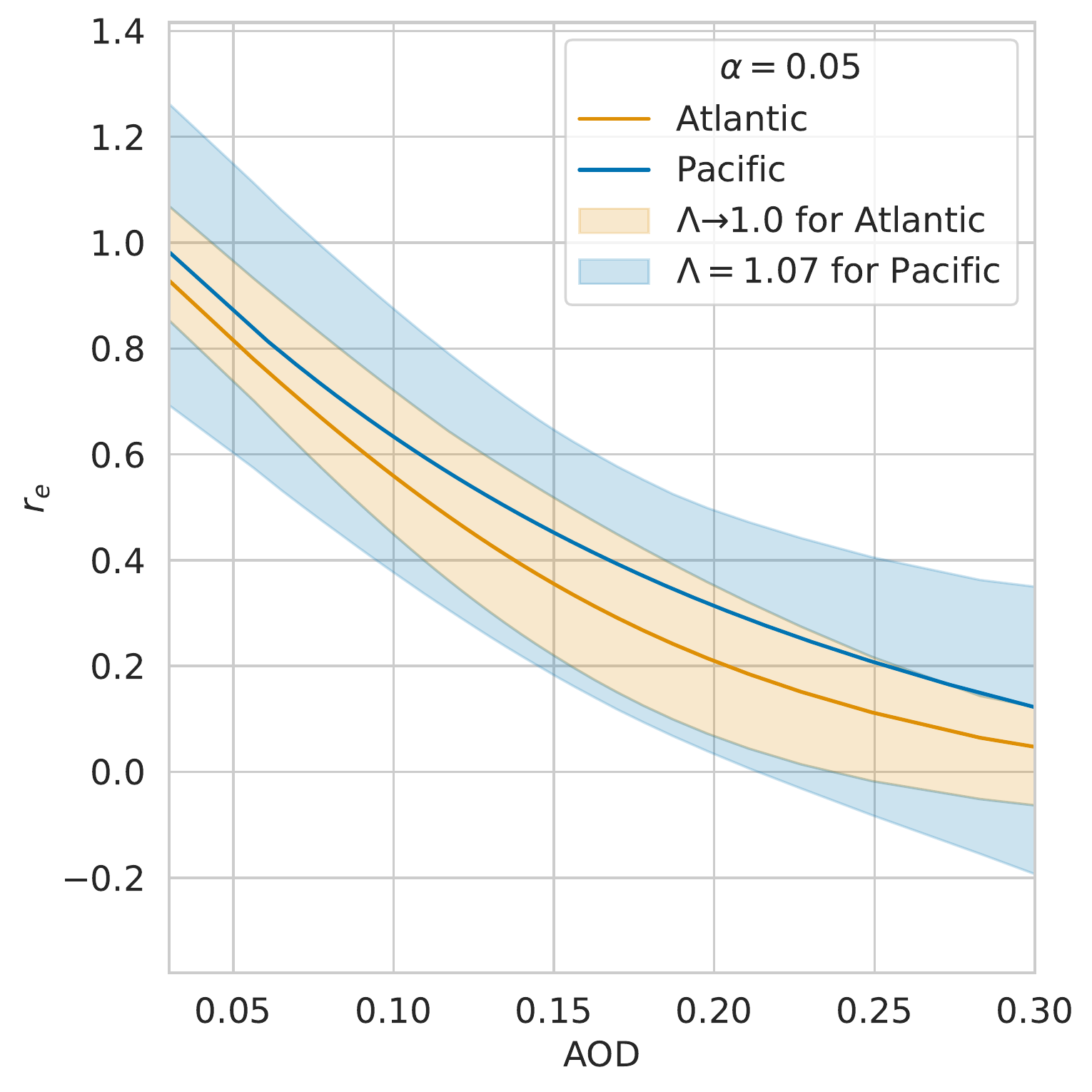}
    \label{fig:lambda}
    }
    \subfigure[Comparison with ESMs in the Pacific]{
    \includegraphics[width=.4\linewidth]{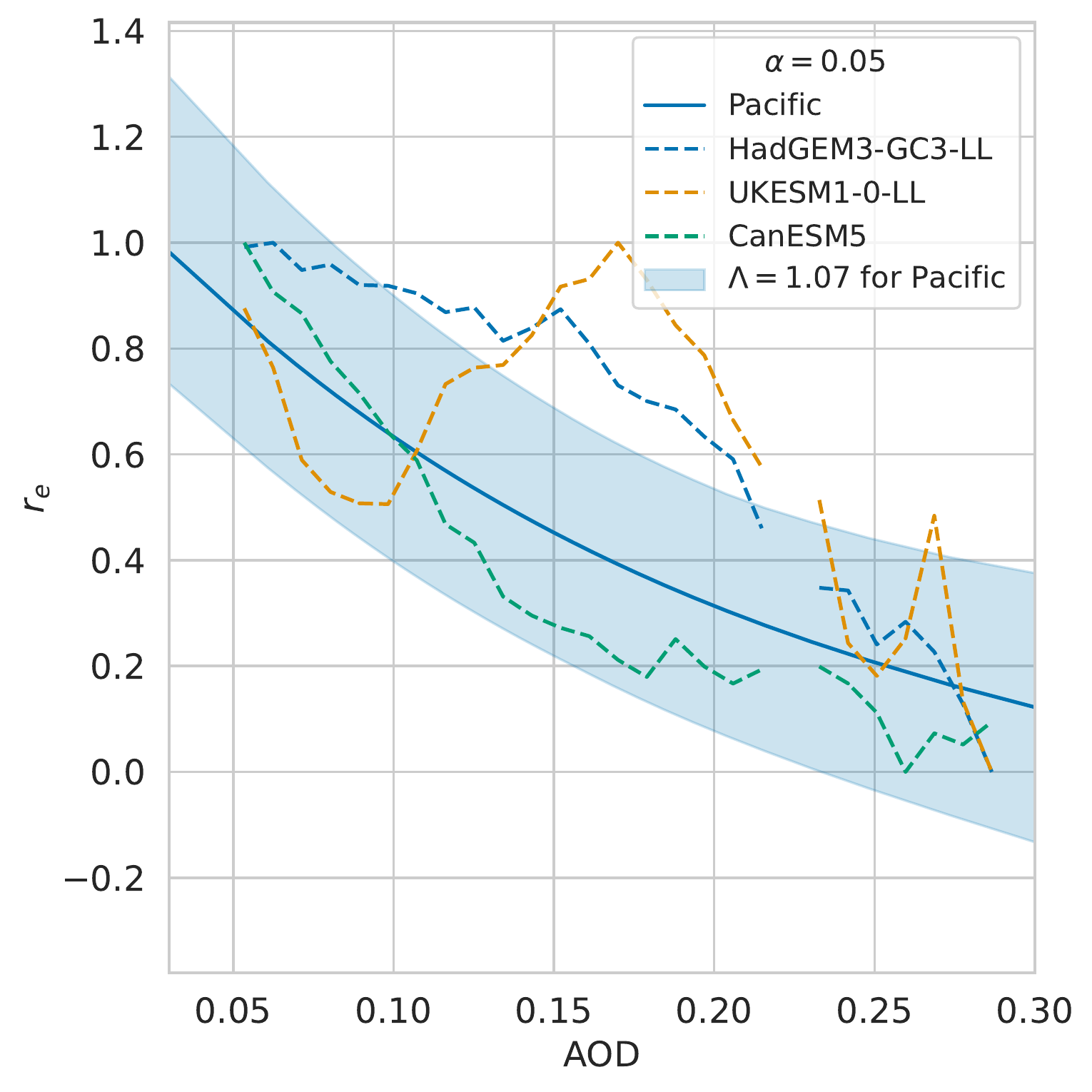}
    \label{fig:esm}
    }
    \caption{Plausible range of effects of aerosol (AOD) on mean droplet radius ($r_e$).}
    \label{fig:results}
\end{figure}

As we now have a possible range of ACI derived using the real, observed outcomes, we can judge how well climate models recreate this observed trend by seeing if their responses lie within our derived interval (\cref{fig:esm}).
We find that the Canadian model CanESM5 simulates ACI better than the UK models HadGEM3-GC3-LL and UKESM1-0-LL.
Our trained machine learning model not only uses the real, observed relationships to derive the magnitude of the effect, but can consider the environmental context and confounding influences to derive real, quantifiable bounds of uncertainty.
Therefore, by using the curves found by Overcast as the true response, we know those models which lie outside of our bounds found by contrasting different regions are likely unphysical and highly unlikely to occur in the real-world.
CanESM5 currently estimates the total cooling effect due to ACI to offset approximately half of the warming due to greenhouse gases; based on our results, we would say it is likely that this estimate is closest to the true value observed on Earth \cite{smith2020effective}.

\section{Discussion and conclusions}

\subsection{Machine learning's place in climate model verification}

In this work, we show that machine learning methods offer viable ways to objectively judge how well global climate models reproduce climate processes such as the effect of aerosol on mean droplet radius.
A drawback of historical studies which utilize satellite observations is their inability to quantify how the surrounding environment may affect the magnitude of the aerosol-cloud interactions.
Overcast accounts for such contextual confounding and communicates bounds on the treatment effect due to an expert-informed influence of hidden-confounding.
Utilizing this method gives us insight into whether climate models reproduce the observed relationships between  AOD and  $r_e$.
Climate models currently only reasonably recreate large scale processes that can be explicitly calculated, leaving processes like aerosol-cloud interactions, which occur on scales smaller than the grid scale, poorly parameterized and approximated.
In order to improve our climate models, we must understand in more relatable terms how well they are doing, such as by comparing their outcomes to those from observations. 
Machine learning provides not only a way to judge these outcomes, but the relationships learned by Overcast and similar models could in the future be fine-tuned to replace our current parameterizations \cite{gettelman2021machine}.

\subsection{Collaboration across domains vs. purely data driven}
\begin{figure}[htbp]
    \centering
    \subfigure[Scaled, with appropriate $\Lambda$]{
    \includegraphics[width=.4\linewidth]{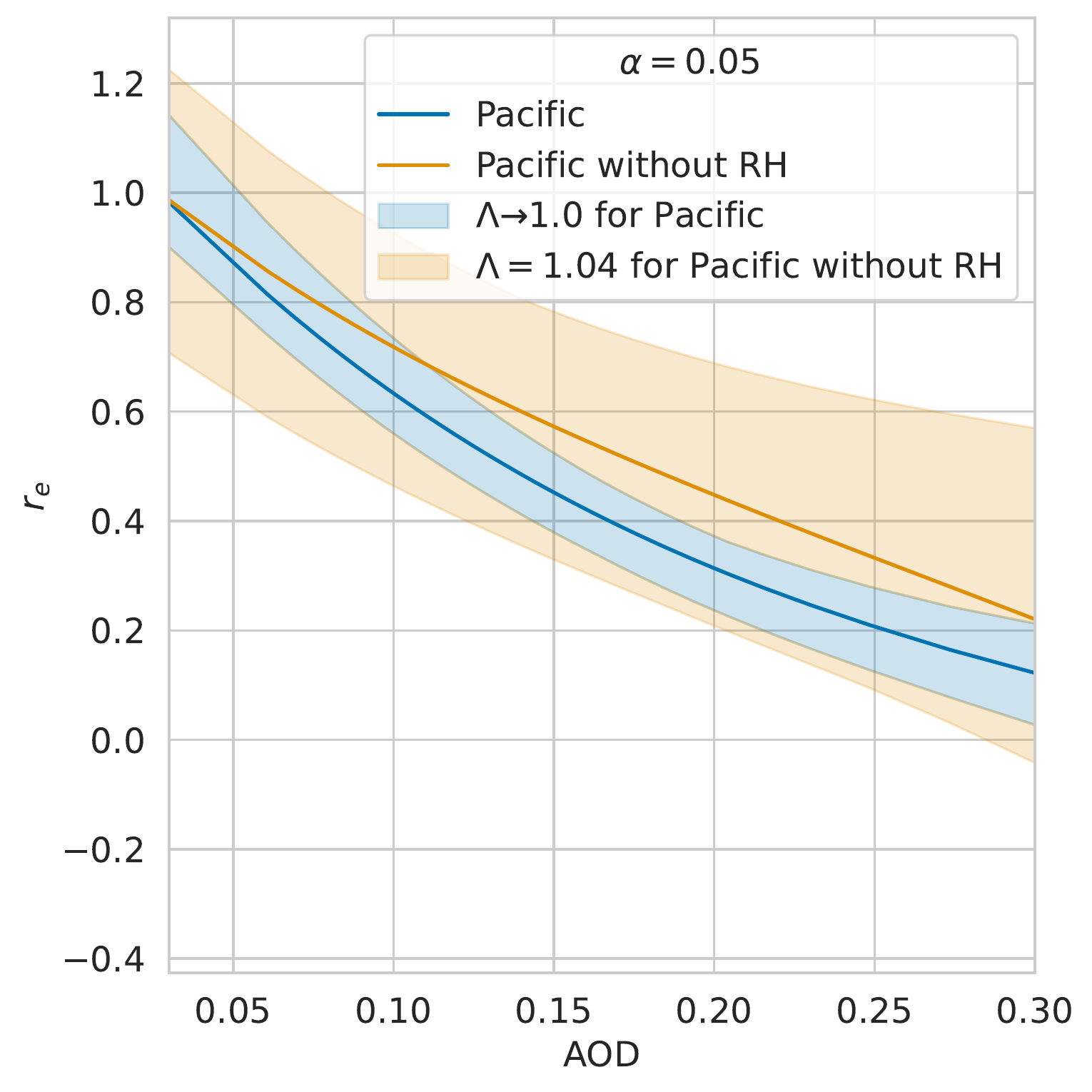}
    \label{subfig:lrprh_scaled_fitted}
    }
    \subfigure[Unscaled]{
    \includegraphics[width=.4\linewidth]{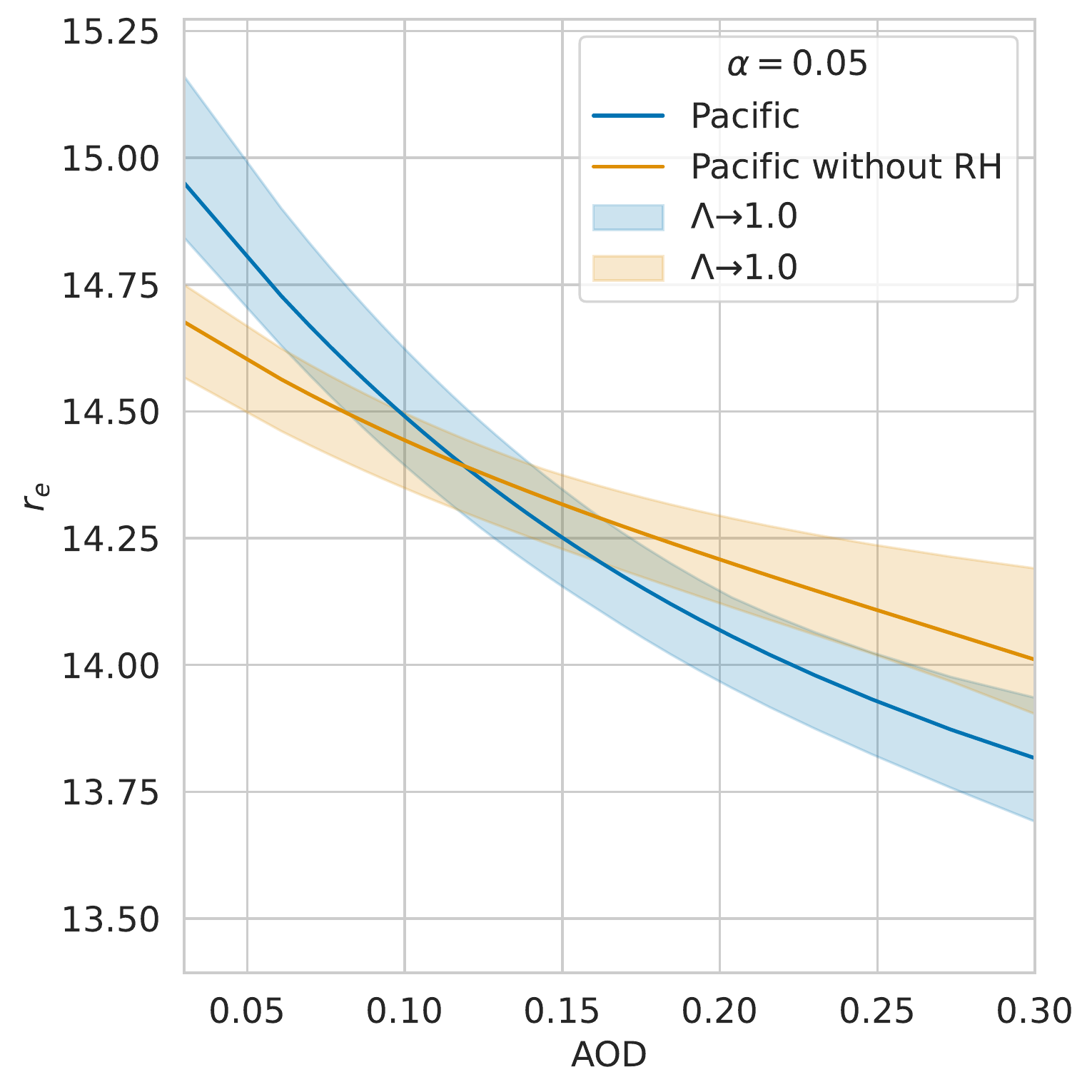}
    \label{subfig:lrprh_unscaled}
    }
    \caption{Plausible range of effects when omitting relative humidity from the covariates.}
    \label{fig:lrprh}
\end{figure}

While different sources of confounding due to regional differences alter the outcomes, the choice of which environmental factors are the main sources of confounding can also be investigated using Overcast.
We perform two experiments, with and without relative humidity at 900, 850 and 700 millibars, to derive varying outcome shapes and fit $\Lambda$ to both dose-response curves.
When $\Lambda$ is set to 1.04, both curves are captured by the bounds of uncertainty, allowing us to view how the response may vary within those bounds due to meteorological uncertainty rather than regional uncertainty, where $\Lambda = 1.07$ was required (\cref{subfig:lrprh_scaled_fitted}).
In the absence of ground truth, purely data-driven techniques cannot decide between the model with and without relative humidity, but as domain knowledge is brought in, it is known that the curve with humidity included is the true response curve (\cref{subfig:lrprh_unscaled}). 
Purely data-driven approaches may not be the most appropriate for studying climatological processes such as aerosol-cloud interactions as domain knowledge is essential to select the correct inputs and verify the outcomes. 
The most robust model arises from combining data and theory, bringing together experts in machine learning and climate processes.

\subsection{Limitations and future works}

This work uses aerosol optical depth as a proxy for aerosol concentration,  which could bias treatment-effect estimates. For example, it is known that bias can arise from measurement-error in the treatment \cite{measurement-error}. Further, we rely on low resolution data that does not perfectly capture the microphysical processes. Future work could consider different assumptions on the underlying causal model, and attempt to include other aerosol properties like size, hygroscopicity and type. 

\section{Acknowledgements}
This project was supported by the European Union’s Horizon 2020 research and innovation program under grant agreement No 821205 (FORCeS), the European Research Council (ERC) project constRaining the EffeCts of Aerosols on Precipitation (RECAP) under the European Union’s Horizon 2020 research and innovation programme with grant agreement no. 724602, and the Turing Institute post-doctoral enrichment award.

\printbibliography

\newpage
\appendix

\section{Theoretical background: unconfoundedness and positivity assumptions}
\label{app:assumptions}

Confounding variables are factors that influence both the treatment $\mathrm{T}$ and the outcomes $\mathrm{Y}$. 
The unconfoundedness assumption states that all confounding variables are observed and controlled for using \(\mathbf{X}\), so that the treatment groups are comparable, that is, \( \mathrm{Y}_\mathrm{T} \indep \mathrm{T} \mid \mathbf{X} \). 

The positivity assumption states that all subgroups of the data with different covariates have a non-zero probability of receiving any dose of treatment, that is, \(p(\mathrm{t} \mid \mathbf{x}) > 0\) for any $\mathrm{t} \in \mathcal{T}$ and for any $\mathbf{x} \in \mathcal{X}$ such that \(p(\mathbf{x}) > 0\).

In practice, there is a trade-off between positivity and unconfoundedness due to the curse of dimensionality, as with large $\mathbf{X}$ and continuous treatment, it is unlikely that we observe all treatment levels for each $\mathbf{x} \in \mathcal{X}$.

\section{Data and pre-processing}
\label{app:data}

We work with data which is retrieved from re-analyses of satellite observations. The Moderate Resolution Imaging Spectroradiometer (MODIS) instruments aboard the Terra and Aqua satellites observe the Earth at approximately 1 km $\times$ 1 km resolution~\cite{baumIntroductionMODISCloud2006}.
These observations are fed into the Modern-Era Retrospective Analysis for Research and Applications version 2 (MERRA-2) real-time model to emulate the atmosphere and its components, such as aerosol~\cite{gelaroModernEraRetrospectiveAnalysis2017}. 
MERRA-2 calculates global vertical profiles of temperature, relative humidity, and pressure, and assimilates hyperspectral and passive microwave satellite observations to enhance its ability to model Earth’s atmosphere. 
The data studied are MODIS observations from the Aqua and Terra satellites collocated with MERRA reanalyses of the environments. 
We work with two different datasets which are $1^\circ \times 1^\circ$ daily means of observations over the South Atlantic and the South East Pacific from 2004 to 2019. 
The sources are given in \cref{tab:data}. 

\begin{table}[htbp]
    \caption{Sources of satellite observations}
    \label{tab:data}
    \centering
    \begin{tabular}{ll}
        \toprule
        Product Name                        & Description                               \\
        \midrule
        Mean Droplet Radius ($r_e$)         & MODIS (1.6, 2.1, 3.7 $\mu$m channels)~\cite{baumIntroductionMODISCloud2006}                                             \\
        Precipitation                       & NOAA CMORPH CDR~\cite{pratGlobalEvaluationGridded2021}\\
        Sea Surface Temperature (SST)       & NOAA WHOI CDR~\cite{coganMeasurementSeaSurface1976}   \\
        Lower Tropospheric Stability (LTS)  & MERRA-2~\cite{gelaroModernEraRetrospectiveAnalysis2017}                                                 \\
        Vertical Motion at 500 mb ($\omega500$)    & MERRA-2~\cite{bosilovichMERRA2InitialEvaluation}      \\
        Estimated Inversion Strength (EIS)  & MERRA-2~\cite{gelaroModernEraRetrospectiveAnalysis2017,woodRelationshipStratiformLow2006}                           \\
        Relative Humidity at x mb (RHx)           & MERRA-2~\cite{gelaroModernEraRetrospectiveAnalysis2017}                                                             \\
        Aerosol Optical Depth (AOD)         & MERRA-2~\cite{gelaroModernEraRetrospectiveAnalysis2017}                       \\
        \bottomrule
    \end{tabular}
\end{table}

We restrict our observations to clouds in the ``aerosol limited'' regime by applying some filtering~\cite{korenaerosollimitedinvigorationwarm2014}.
In ``aerosol limited'' regimes, we assume that cloud development is limited by the availability of cloud-condensation nuclei, and thus aerosol. 
Our choice of filtering is informed by domain knowledge.
CWP are filtered to values below 250$\mu$m and $r_e$ to values below 30$\mu$m. 
AOD values are filtered, only keeping values between 0.03 and 0.3. 
We also filter out precipitating clouds to avoid a loop in the causal graph.
Finally, all features are normalized before being fed into the model. 

\section{Model architecture}
\label{app:model-arch}

\begin{figure}[htbp]
    \centering
    \includegraphics[width=\linewidth]{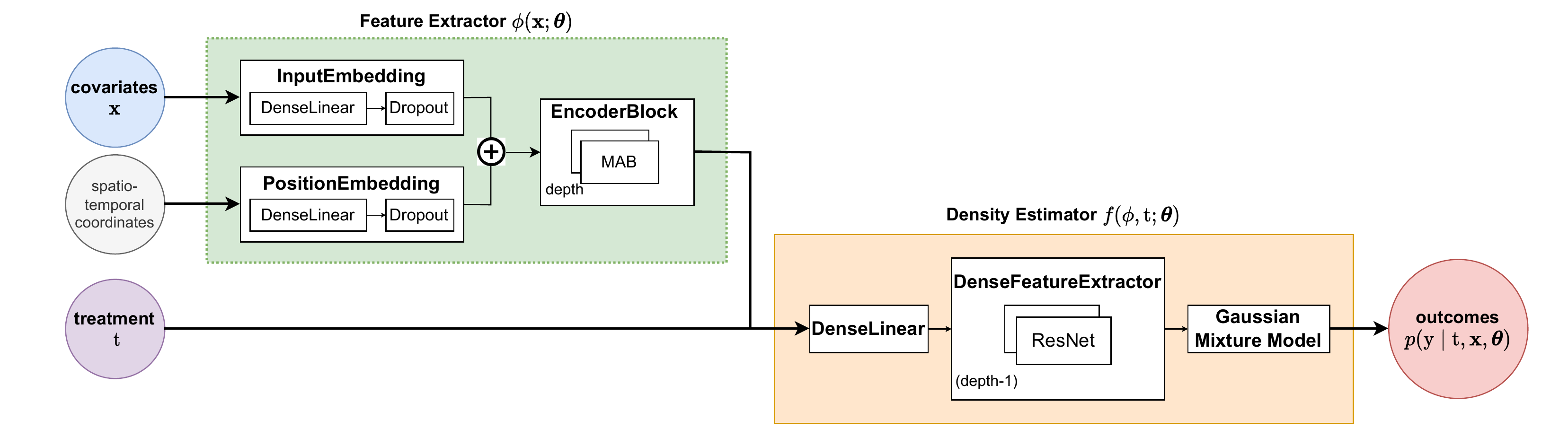}
    \caption[Overcast model architecture]{\textbf{Overcast model architecture.} The inputs are represented by circles, in \textcolor{blue}{\textbf{blue}} the covariates, in \textcolor{grey}{grey} the spatio-temporal coordinates, in \textcolor{purple}{\textbf{purple}} the treatment. In the \textcolor{red}{\textbf{red}} circle is the output of the model, the outcomes distribution. The model has a feature extractor in \textcolor{green}{green} and a density estimator (in \textcolor{orange}{orange}).}
    \label{fig:overcast}
\end{figure}

The models are neural-network architectures with two components: a feature extractor $\phi(\mathrm{x}; \mathbf{\theta})$ and a density estimator $f(\phi, \mathrm{t}; {\theta})$, \added[id=MS]{represented in \cref{app:model-arch}.}
The covariates $\mathbf{x}$ are given as input to the feature extractor, whose output is concatenated with the treatment $\mathrm{t}$ and given as input to the density estimator which outputs a Gaussian mixture density, $p(\mathrm{y} \mid \mathrm{t}, \mathbf{x}, {\theta})$, from which we can sample.
The feature extractor uses attention mechanisms to model the spatio-temporal correlations between the covariates on a given day using the geographical coordinates of the observations.
The model architecture is represented in \cref{fig:overcast}. 

\section{Implementation details}
\label{app:implementation}

We follow the implementation from~\cite{jesson2022scalable}.
The code is written in python.
The packages used include PyTorch~\cite{pytorch}, scikit-learn~\cite{scikit-learn}, Ray~\cite{moritzRayDistributedFramework2018}, NumPy, SciPy and Matplotlib.

We use ray tune~\cite{liawTuneResearchPlatform2018} with HyperBand Bayesian Optimization~\cite{falknerCombiningHyperbandBayesian} search algorithm to optimise our network hyper-parameters.
The hyper-parameters considered during tuning are given in \cite{jesson2022scalable}.
The final hyper-parameters for each dataset are given in \cref{table:hyperparams_final}.
The hyper-parameter optimization objective is the batch-wise Pearson correlation averaged across all outcomes on the validation data for a single dataset realization with random seed 1331.

We split the data into training, validation, and testing sets across different days.
\cite{jesson2022scalable} splits data in the following way: datapoints from Mondays to Fridays are in the training set, from Saturdays in the validation set, and from Sundays in the testing set. 
In our implementation, we keep the same ratio between datasets but we randomize the splits, using random seed 42 and having $5/7$ of the data in the training set, $1/7$ in the validation set, and $1/7$ in the testing set. 
The randomization is motivated by the fact that there is a clear weekly cycle of aerosol optical depth \cite{christensenOpportunisticexperimentsconstrain2022}. 
Models are optimized by maximizing the log likelihood of $p(\mathrm{y} \mid \mathrm{t}, \mathbf{x}, \theta)$.

\begin{table}[htbp]
\caption{Final hyper-parameters for each dataset and model}
\label{table:hyperparams_final}
\centering
\begin{tabular}{lcc}
\hline
\multicolumn{1}{c}{Hyper-parameter} & South-East Pacific & South Atlantic \\ \hline
Hidden Units                        & 128                & 128            \\
Network Depth                       & 3                  & 4              \\
GMM $\mathrm{T}$ Components                 & 27                 & 7              \\
GMM $\mathrm{Y}$ Components                 & 22                 & 24             \\
Attention Heads                     & 8                  & 8              \\
Negative Slope                      & 0.28               & 0.19           \\
Dropout Rate                        & 0.42               & 0.16           \\
Layer Norm                          & False              & True           \\
Batch Size                          & 128                & 160            \\
Learning Rate                       & 0.0001             & 0.0001         \\
Epochs                              & 500                & 500            \\ \hline
\end{tabular}
\end{table}

\end{document}